\documentclass[11pt,a4paper]{article}
\pdfoutput=1
\usepackage{jheppub}
\usepackage{rotating}
\usepackage{array}
\usepackage{amsmath}
\usepackage[normalem]{ulem}
\usepackage{slashed}
\usepackage{booktabs}
\usepackage[pdftex,table]{xcolor}
\usepackage{units}

\usepackage{multirow}

\preprint{DESY-17-016}

\title{Hunting the dark Higgs}

\author[a]{Michael Duerr,}
\author[a]{Alexander Grohsjean,}
\author[a]{Felix Kahlhoefer,}
\author[b]{Bjoern Penning,}
\author[a]{\mbox{Kai Schmidt-Hoberg}}
\author[a]{and Christian Schwanenberger}

\affiliation[a]{DESY, Notkestra\ss e 85, D-22607 Hamburg, Germany}
\affiliation[b]{University of Bristol, HH Wills Physics Laboratory, Tyndall Avenue, Bristol BS8 1TL, UK}

\emailAdd{michael.duerr@desy.de}
\emailAdd{alexander.grohsjean@desy.de}
\emailAdd{felix.kahlhoefer@desy.de}
\emailAdd{penning@cern.ch}
\emailAdd{kai.schmidt-hoberg@desy.de}
\emailAdd{christian.schwanenberger@desy.de}

\abstract{We discuss a novel signature of dark matter production at the LHC resulting from the emission of an additional Higgs boson in the dark sector. The presence of such a dark Higgs boson is motivated simultaneously by the need to generate the masses of the particles in the dark sector and the possibility to relax constraints from the dark matter relic abundance by opening up a new annihilation channel.  If the dark Higgs boson decays into Standard Model states via a small mixing with the Standard Model Higgs boson, one obtains characteristic large-radius jets in association with missing transverse momentum that can be used to efficiently discriminate signal from backgrounds. We present the sensitivities achievable in LHC searches for dark Higgs bosons with already collected data and demonstrate that such searches can probe large regions of parameter space that are inaccessible to conventional mono-jet or di-jet searches.}


\begin{document}
\maketitle

\section{Introduction}

There is considerable interest in the idea that the dark matter~(DM) particle interacts with Standard Model~(SM) states via the exchange of one or more new mediators~\cite{Frandsen:2012rk,Alves:2013tqa,
Arcadi:2013qia, Garny:2014waa, Chala:2015ama, Alves:2015mua,Ghorbani:2015baa,
Jacques:2016dqz, Bell:2016fqf,Duerr:2016tmh,Ko:2016ybp,Goncalves:2016iyg,Bell:2016ekl}, which can for example carry spin 1 (e.g.\ a new $Z'$ gauge boson) or spin 0 (e.g.\ an additional Higgs boson). The presence of such new mediators can lead to observable signals in a wide range of DM searches, in particular direct~\cite{Tan:2016zwf,Akerib:2016vxi, Amole:2015pla,Amole:2016pye} and indirect detection experiments~\cite{Ackermann:2015zua,Aguilar:2016kjl,Aartsen:2016exj} and searches for missing transverse momentum at the Large Hadron Collider~(LHC)~\cite{Buchmueller:2014yoa, Fairbairn:2014aqa,Harris:2015kda,Alves:2015pea, Choudhury:2015lha,Blennow:2015gta, Heisig:2015ira,Buchmueller:2015eea,Brooke:2016vlw}. These mediators could also be responsible for establishing thermal equilibrium between the visible and the dark sector in the early Universe and provide the annihilation and creation processes that set the DM relic abundance via thermal freeze-out~\cite{Busoni:2014gta,Chala:2015ama,Jacques:2016dqz}.

Experimental results, in particular bounds on new resonances and measurements of Higgs production and decay at the LHC~\cite{Khachatryan:2016vau}, strongly constrain the parameter space in which the DM particles can obtain their relic abundance from direct annihilation into SM final states~\cite{Duerr:2016tmh}. This tension can be significantly relaxed if the DM particle is not the lightest state in the dark sector, leading to new annihilation channels. 
Such a new state arises naturally if the DM mass is generated via a Higgs mechanism in the dark sector. The resulting dark Higgs boson $s$ can be lighter than the DM particle $\chi$, so that the DM relic abundance is dominantly set by the process $\chi \chi \to s s$, followed by decays of $s$ into SM states.\footnote{We emphasise that in this paper $s$ always refers to the dark Higgs boson and not to the strange quark.} In this case the observed relic abundance can be readily reproduced~\cite{Kahlhoefer:2015bea,Bell:2016uhg,Bell:2016fqf}.

The relic density then depends only on the coupling $y_\chi$ between the DM particle and the dark Higgs boson, and the couplings to SM particles can be very small. The dark sector can then be highly secluded from the SM~\cite{Pospelov:2007mp} and will be very difficult to probe using conventional DM searches even if the coupling between DM and the dark Higgs boson is large~\cite{LopezHonorez:2012kv}. Moreover, the DM annihilation rate into dark Higgs bosons is velocity suppressed and therefore unobservable in the present Universe~\cite{Bell:2016uhg,Duerr:2016tmh}. Astrophysical constraints from DM self-interactions are also not sensitive to this scenario unless the mass of the dark Higgs boson is several orders of magnitude below the mass of the DM particle. 

A promising way to probe these models opens up if there is another mechanism to produce dark sector states at the LHC (for example via an additional $Z'$ mediator), because any such state can radiate off a dark Higgs boson. Since the couplings within the dark sector are typically large in order to reproduce the observed relic abundance, the probability of dark-Higgs strahlung can be large despite the very small couplings of the dark Higgs boson to the SM. If the dark Higgs boson is the lightest state in the dark sector it further decays into SM particles. The emission of a visibly decaying dark Higgs boson then indicates the production of DM.

\begin{figure}[tb]
\centering
\includegraphics[height=0.15\textheight,clip,trim = 0 650 430 0]{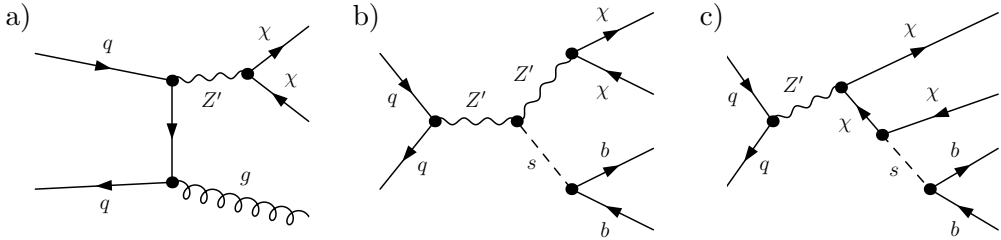}\qquad
\includegraphics[height=0.15\textheight,clip,trim = 190 650 260 0]{Diagrams.pdf}\qquad
\includegraphics[height=0.15\textheight,clip,trim = 360 650 60 0]{Diagrams.pdf}
\put(-400,85){a)}
\put(-270,85){b)}
\put(-140,85){c)}
\caption{Processes leading to missing transverse momentum signatures at the LHC. Left: a typical mono-jet process. Centre and right: processes leading to a mono-dark-Higgs signal.}
\label{fig:diagrams}
\end{figure} 

In contrast to conventional DM signatures at the LHC, where missing transverse momentum results from the recoil of DM particles against a SM state from initial state radiation (see figure~\ref{fig:diagrams}a), here the DM particles recoil against a visibly decaying dark Higgs boson from final state radiation (figure~\ref{fig:diagrams}b, \ref{fig:diagrams}c). Compared to, for example, mono-jet searches, the crucial difference is that in most of the interesting parameter space the dark Higgs boson will decay into a pair of heavy quarks, leading to a characteristic highly boosted large-radius jet (a fat jet) that can be readily distinguished from many background processes.

A similar experimental signature is considered in the context of so-called mono-Higgs searches~\cite{Petrov:2013nia,Carpenter:2013xra,Berlin:2014cfa}, making use of highly refined tagging algorithms for Higgs bosons with high transverse momentum~($p_\mathrm{T}$)~\cite{Aad:2015dva,Aaboud:2016obm}. The probability to produce a SM Higgs boson in association with DM is however rather small if the mixing between the dark Higgs boson and the SM Higgs boson is small (as required by the observed properties of the SM Higgs boson~\cite{Khachatryan:2016vau}).\footnote{A possible exception are DM models based on two Higgs doublet models, where the mediator mixes with the neutral component of the second Higgs doublet rather than with the SM-like Higgs boson~\cite{Ipek:2014gua,No:2015xqa,Goncalves:2016iyg,Bauer:2017ota}.} Dark-Higgs strahlung, on the other hand, can be large without modifying the properties of SM Higgs boson at all.

It is therefore highly promising to apply the strategies developed in the context of mono-Higgs searches to search for additional dark Higgs bosons. In this paper we propose mono-dark-Higgs searches as a new way to probe dark sectors. We study how existing methods for tagging fat jets from the decay of highly boosted Higgs bosons can be applied to additional dark Higgs bosons. The resulting searches can be sensitive to interesting regions of parameter space that are currently inaccessible for most missing transverse momentum searches at the LHC.

The paper is structured as follows. In section~\ref{sec:framework} we introduce a simple framework for a dark Higgs boson and discuss the connection to existing DM models. We present the motivation for mono-dark-Higgs searches and a first estimate of their potential sensitivity. Section~\ref{sec:backgrounds} then details the estimation of SM backgrounds for these searches. We discuss the possibility to identify jets resulting from the decay of a boosted dark Higgs boson and how this can be used to suppress backgrounds. The calculation of the expected signal is given in section~\ref{sec:sensitivity} together with our estimate of the sensitivity of possible LHC searches. In section~\ref{sec:relic} we combine our results with a calculation of the DM relic abundance to point out the remarkable complementarity between mono-dark-Higgs searches and searches for di-jet resonances. Our conclusions are presented in section~\ref{sec:conclusions}.

\section{Framework for a dark Higgs boson}
\label{sec:framework}

We consider a Majorana DM particle $\chi$ that obtains its mass from the vacuum expectation value~(vev) $w$ of a new complex Higgs field $S$, which is a singlet under the SM gauge group~\cite{Kahlhoefer:2015bea,Duerr:2016tmh}. The field $S$ carries a charge $q_S$ under a new $U(1)'$ gauge group, so its vev $w$ breaks the gauge symmetry spontaneously and generates the mass of the corresponding $Z'$ gauge boson. The symmetry breaking gives rise to a new physical Higgs boson $s$, defined via $S = 1/\sqrt{2} ( s + w )$, which we will call the dark Higgs boson. In this set-up, the DM particle acquires an axial coupling to the $Z'$, so that all three particles in the dark sector couple to each other. The renormalisable terms in the Lagrangian are given by:
\begin{equation}
 \mathcal{L}_\chi = - \frac{1}{2} g^\prime q_\chi Z^{\prime \mu} \bar{\chi} \gamma^5 \gamma_\mu \chi - \frac{y_\chi}{2 \sqrt{2}} s \bar{\chi} \chi + \frac{1}{2} g^{\prime 2} q_S^2 Z^{\prime \mu} Z^\prime_\mu \left( s^2 + 2 s w \right) \; , 
\end{equation}
where $g^{\prime}$ denotes the $U(1)'$ gauge coupling and gauge invariance requires that the charge of the DM particle satisfy $q_\chi = q_S / 2$. 

The masses of the DM particle and the $Z'$ are respectively given by $m_\chi = y_\chi \, w / \sqrt{2}$ and \mbox{$m_{Z'} = 2 \, g' \, q_\chi \, w$}, while the mass of the dark Higgs boson $m_s$ is an independent parameter, which should not be much larger than $w$ in order to preserve perturbative unitarity~\cite{Kahlhoefer:2015bea,Duerr:2016tmh}. The dark sector thus contains four independent parameters, which we take to be $m_\chi$, $m_{Z'}$, $m_s$ and $g_\chi \equiv g' \, q_\chi$.\footnote{An analogous discussion applies in the case that the DM particle is a Dirac fermion. The only difference is that in this case DM can also have a vectorial coupling $g_V$ to the $Z'$, which is independent of all other parameters. For simplicity we focus here on the more constrained scenario with only four free parameters.} In terms of these parameters the interaction Lagrangian yields:
\begin{equation}
 \mathcal{L}_\chi = - \frac{1}{2} g_\chi Z^{\prime \mu} \bar{\chi} \gamma^5 \gamma_\mu \chi - g_\chi \frac{m_\chi}{m_{Z'}} s \bar{\chi} \chi + 2 \, g_\chi \, Z^{\prime \mu} Z^\prime_\mu \left( g_\chi \, s^2 + m_{Z'} s \right) \; .
\label{eq:Ldark}
\end{equation} 

We will be interested in the case where the DM particle is not the lightest state in the dark sector, so that it can annihilate into other dark sector states which subsequently decay into SM states.  Scenarios in which the $Z'$ is the lightest particle are typically disfavoured, because they require the couplings of the dark Higgs boson to be close to the perturbativity bound.\footnote{For a discussion of mono-$Z'$ searches, we refer to~\cite{Bai:2015nfa, Autran:2015mfa}.} We therefore focus on the more natural case that the dark Higgs boson is the lightest state in the dark sector and the relic density is largely set by the process $\chi \chi \rightarrow s s$.

The two mediators offer three different possibilities in which such a dark sector can be coupled to the SM: via direct couplings of the $Z'$ to SM particles, via mixing of the $Z'$ with the neutral gauge bosons of the SM or via mixing between the dark Higgs boson and the SM Higgs boson~\cite{Frandsen:2012rk,Kahlhoefer:2015bea}. In particular, non-zero mixing between the dark Higgs boson and the SM Higgs boson ensures that the dark Higgs boson is unstable even if it is the lightest state in the dark sector and decays into SM states with a negligible lifetime. The required mixing angle can however be so small that it does not lead to any other observable effects. For the purpose of this work we assume that the dominant interaction results from vector couplings of the $Z'$ to quarks, which naturally arise in models of gauged baryon number~\cite{Pais:1973mi,FileviezPerez:2010gw,Liu:2011dh,Duerr:2013dza,Duerr:2013lka,Perez:2014qfa,Duerr:2014wra,Duerr:2015vna,Ohmer:2015lxa}:
\begin{equation}
  \mathcal{L}_\chi = - g_q Z^{\prime \mu} \bar{q} \gamma_\mu q \; .
\label{eq:Lquark}
\end{equation}

Axial-vector couplings of the $Z^\prime$ to the SM quarks would also generate the signal we consider. However, such couplings require a modification of the set-up to guarantee SM gauge invariance~\cite{Kahlhoefer:2015bea} and will therefore not be discussed for simplicity. Additional couplings of the $Z^\prime$ to SM leptons could also be present and will only have a negligible effect on the signal. Such couplings will lead to additional constraints~\cite{Ekstedt:2016wyi} and we do not discuss this possibility further. Thus, the scenario we consider is a simplified version of the models for DM from gauged baryon number~\cite{Duerr:2013lka,Duerr:2013dza,Duerr:2014wra,Perez:2014qfa,Duerr:2015vna,Ohmer:2015lxa}, where we do not specify the additional fermion content necessary for anomaly cancellation. 

There is also an obvious similarity to the spin-1 simplified DM model studied by the LHC collaborations~\cite{Buchmueller:2013dya,
Harris:2014hga, Buckley:2014fba,
Abdallah:2015ter,Abercrombie:2015wmb,Kahlhoefer:2015bea,
Englert:2016joy,Boveia:2016mrp}, with the one addition that we specify the mechanism responsible for generating the masses of the DM particle and the $Z'$ and for this purpose introduce a dark Higgs.\footnote{Another more subtle difference is that we consider a Majorana DM particle. For this case the vector coupling of DM to the $Z'$ is forbidden and the sensitivity of direct detection experiments is strongly suppressed.} Since the couplings of the dark Higgs boson are fully specified by the other parameters in the model, the only new parameter that we introduce is the dark Higgs mass $m_s$ (the precise value of the dark Higgs mixing angle is irrelevant for the phenomenology). To facilitate the comparison with existing LHC searches we adopt the same coupling choice used by the LHC collaborations and focus on $g_q = 0.25$ and \mbox{$g_\chi = 1$}~\cite{Abercrombie:2015wmb}. We note that for these couplings the observed DM relic abundance will only be reproduced for specific values of the three masses. We will return to this issue in section~\ref{sec:relic}.

While the presence of a light dark Higgs boson can help to avoid DM overproduction in the early Universe and therefore relax cosmological constraints on the parameter space of the model, it will also lead to new constraints from the LHC. To understand the origin of these constraints let us consider the Drell--Yan production of a $Z'$ and consider the possible decay modes. In the absence of a light dark Higgs boson, the $Z'$ decays either into SM quarks or (provided $m_{Z'} > 2 m_\chi$) into a pair of DM particles, with the branching ratios depending on the ratio of $g_q^2 / g_\chi^2$. In the presence of a dark Higgs boson, a third decay mode becomes available, namely the three-body decay $Z' \rightarrow \chi \chi s$ (see figure~\ref{fig:diagrams}b,~\ref{fig:diagrams}c). Although this decay mode requires an off-shell intermediate state (either a $Z'$ or a DM particle) and is phase-space suppressed, the branching ratio is not negligible. For $g_q = 0.25$ and $g_\chi = 1$ the three-body decay involving a dark Higgs boson can have a branching fraction of more than $5\%$, provided $m_{Z'} \gg m_\chi, \, m_s$. 

In a large fraction of these three-body decays the dark Higgs boson will be relatively soft and the two DM particles will be approximately back-to-back. Therefore only little missing transverse momentum ($\slashed{E}_\mathrm{T}$) is produced in the decay. Nevertheless, there is a sizeable probability that the dark Higgs boson has a large momentum. For $m_{Z'} = \unit[2]{TeV}$ (and $m_\chi, \, m_s \ll m_{Z'}$), in approximately 10\% of the three-body decays the dark Higgs boson will have an energy exceeding $\unit[500]{GeV}$. After its decay into SM particles one obtains one or several high-$p_\mathrm{T}$ jets in association with large missing transverse momentum, a very powerful signature in LHC searches. 

\begin{figure}[tb]
\centering
\includegraphics[width=0.5\textwidth]{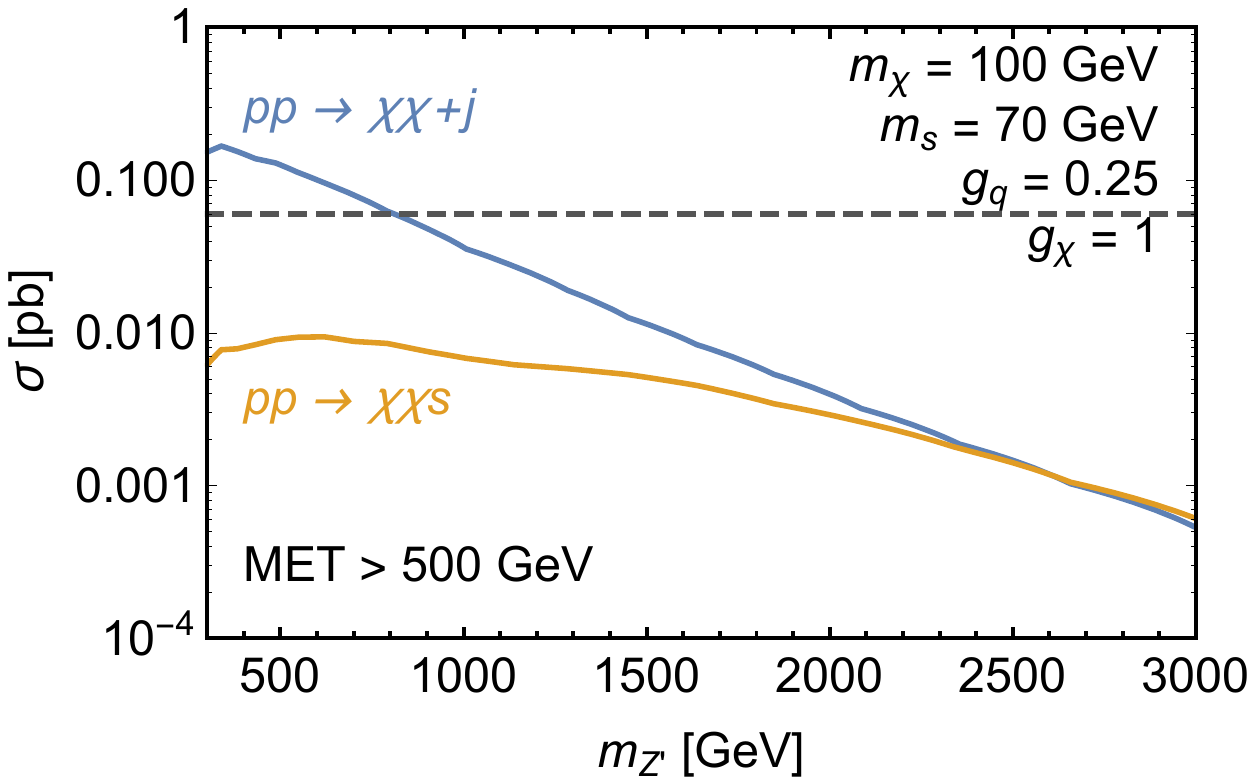}
\caption{Total cross section for $p p \rightarrow \chi \chi + j$ and $p p \rightarrow \chi \chi s$ as a function of the $Z'$ mass for a specific choice of the remaining masses and couplings. To calculate the cross section, we require $\slashed{E}_\mathrm{T} > \unit[500]{GeV}$ after showering and impose the selection cuts from the most recent ATLAS mono-jet search~\cite{Aaboud:2016tnv} (see text for details). The dashed line indicates the corresponding bound on the cross section obtained by ATLAS.}
\label{fig:MET}
\end{figure} 

Such high-$p_\mathrm{T}$ jets can also be produced from initial state radiation: $p p \rightarrow Z' + j \rightarrow \chi \chi + j$. However, if the $Z'$ is too heavy to be produced together with additional energetic states, the most efficient way to obtain large missing transverse momentum may be from the decay of $Z'$ bosons produced roughly at rest. This is illustrated in figure~\ref{fig:MET}, which shows a comparison of the cross section of a conventional mono-jet signal $pp \rightarrow \chi\chi + j$ (blue line) and a mono-dark-Higgs signal $pp \rightarrow \chi\chi s$ (orange line) as a function of the $Z'$ mass.\footnote{In both processes, we allow for an additional jet when producing the signal samples. The details of the signal generation are discussed in section~\ref{sec:sensitivity}} In both cases we require $\slashed{E}_\mathrm{T} > \unit[500]{GeV}$ after showering  and apply the selection cuts from the most recent ATLAS mono-jet search~\cite{Aaboud:2016tnv}. Specifically, we require $p_\mathrm{T} > \unit[250]{GeV}$ for the leading jet, not more than four jets with $p_\mathrm{T} > \unit[30]{GeV}$ and $|\eta| < 2.8$, and veto events with electrons (muons) with $p_\mathrm{T} > \unit[20]{GeV}$ ($p_\mathrm{T} > \unit[10]{GeV}$) and $|\eta| < 2.47$ ($|\eta| < 2.5$). Moreover, events are also rejected if any of the jets point in the same azimuthal direction as the missing energy vector, i.e.\ if $\Delta\phi(j,\slashed{E}_\mathrm{T}) < 0.4$.

For small $Z'$ masses large amounts of missing transverse momentum can only be generated if either the $Z'$ is produced off-shell or in association with a jet from initial state radiation. The conventional mono-jet signal then benefits from the larger branching ratio compared to the mono-dark-Higgs signal. For a heavy $Z'$ however large missing transverse momentum can be generated in the three-body decay without requiring an additional hard jet. As a result, the total cross section for the mono-dark-Higgs signal decreases much more slowly with increasing $m_{Z'}$ and becomes comparable to the mono-jet signal for $m_{Z'} \gtrsim \unit[2]{TeV}$.

The dashed line in figure~\ref{fig:MET} indicates the present sensitivity of LHC mono-jet searches (the bounds from~\cite{Aaboud:2016tnv} on the cross section of DM events with $\slashed{E}_\mathrm{T} > \unit[500]{GeV}$ are used). 
We conclude that searches for standard jets in association with missing transverse momentum are presently not sensitive to $m_{Z'} > \unit[1]{TeV}$ for the couplings assumed. However, it is possible to experimentally distinguish the jets produced from the decay of a boosted dark Higgs boson from ordinary jets. This is because the dark Higgs boson obtains its couplings from mixing with the SM Higgs boson and will therefore decay dominantly into heavy quarks (assuming $2 \, m_b \sim \unit[10]{GeV} < m_s <  \unit[160]{GeV} \sim 2 \, m_W$), resulting in a single fat jet containing two $b$-jets.\footnote{A similar experimental signature has also been considered in the context of exotic Higgs decays~\cite{Huang:2013ima,Huang:2014cla}.} As we will see, for this specific signature SM backgrounds are highly suppressed. This enables to achieve much better sensitivities compared to common mono-X searches, even if the cross section for $p p \rightarrow Z' \rightarrow \chi \chi s$ is smaller than the one for $p p \rightarrow Z' + X \rightarrow \chi \chi + X$.

We note that the dark Higgs production cross section may be sizeable even if $2 \, m_\chi + m_s > m_{Z'}$.  This is because for $m_{Z'} > 2 \, m_\chi$ the process shown in figure~\ref{fig:diagrams}b can still take place but with a $Z'$ that is off-shell in the first step and on-shell in the second step: $p p \rightarrow Z'^\ast \rightarrow Z' s \rightarrow \chi\chi s$. This way the mono-dark-Higgs signal can still be resonantly enhanced.

\section{Dark Higgs tagging and background estimates}
\label{sec:backgrounds}

\subsection{Tagging a dark Higgs}

The expected experimental signature is a scalar resonance decaying into a $b\bar{b}$-pair produced in association with large amounts of missing transverse momentum. The signature is however quite different from the one obtained in models where a DM pair is radiated from a bottom quark produced in a QCD process~\cite{Aad:2014vea}. Because the scalar resonance is typically light and highly boosted, the decay products are collimated and the $b\bar{b}$-pair merges into a single fat jet with an invariant mass corresponding to the mass of the scalar resonance. Hence, the search for a dark Higgs boson is to first approximation the search for a peak in the invariant mass distribution of fat jets produced in association with missing transverse momentum.

Nevertheless, a number of techniques to further refine the identification of so-called Higgs jets and reject potential backgrounds have been developed in existing searches for SM Higgs bosons with high-$p_\mathrm{T}$~\cite{ATLASanalysis}. A first step is to apply $b$-tagging techniques to the fat jet and consider only the invariant mass of $b$-tagged fat jets. Further improvements are however possible by considering the substructure of the fat jet. For example, ref.~\cite{ATLASperformance} uses an approach in which a fat jet is tagged as a Higgs jet if it contains two $b$-tagged track jets with a smaller radius parameter. While ref.~\cite{ATLASperformance} focusses on the tagging of fat jets from SM Higgs decays, a similar approach can be developed for a dark Higgs boson with a mass different from that of the SM Higgs boson. We implement such an analysis in \texttt{Rivet~v2.5.2}~\cite{Buckley:2010ar} as follows.

Fat jets are reconstructed from the truth particles except for muons and neutrinos using \texttt{FastJet~v3.2.0}~\cite{Cacciari:2011ma} and an anti-$k_t$ algorithm~\cite{Cacciari:2008gp} with a distance parameter of $R=1.0$. A trimming algorithm~\cite{Krohn:2009th} discarding the softer components of the fat jets is applied. For this purpose, $k_t$ subjets with a distance parameter of 0.2 are built and subjets are removed if their transverse momentum is less than 5\% of the total transverse momentum of the large-$R$ jet. After trimming, jets are required to have a transverse momentum $p_{\mathrm T} > \unit[250]{GeV}$ and a pseudo-rapidity $|\eta| < 2.0 $. 

To identify fat jets from a dark Higgs decay to a pair of bottom quarks, the flavour of small track jets associated to the fat jet is used. To do so, track jets are reconstructed from all charged truth final state particles employing an anti-$k_t$ algorithm with $R=0.2$. Track jets are required to have at least two tracks, $p_{\mathrm T} > \unit[10]{GeV}$ and \mbox{$|\eta| < 2.5 $}. The flavour of each track jet is estimated using $b$-hadrons and the ghost-association technique~\cite{Cacciari:2008gn}. An average efficiency of 70\% is assumed for the proper identification of jets containing $b$-hadrons. A misidentification probability of 12\% is used for jets containing $c$-hadrons and of 0.6\% for light-flavour jets~\cite{ATLASanalysis}. Fat jets are required to be geometrically matched to at least two $b$-hadron tagged track jets where the maximum distance between the fat jet and the track jet should be less than 1.1. To account for muons inside a fat jet, the four-momentum of the fat jet is corrected for the missing four-momentum of the closest muon within $\Delta R <  0.2$ to its associated track jets. In a final step, the jet mass is smeared assuming a 10\% mass resolution, as observed in ref.~\cite{ATLASperformance}.              

Besides containing exactly one good fat jet, selected events are required to have a minimum transverse energy of $\unit[500]{GeV}$, where the missing transverse momentum is reconstructed as the negative sum of the four-momenta of all visible final states and smeared according the resolution observed by the ATLAS experiment~\cite{Aad:2012re}. Moreover, events are vetoed if they contain at least one prompt, isolated lepton with $p_{\mathrm T} > \unit[7]{GeV}$ and $| \eta | < 2.5$. 

We validate our analysis as follows. The separation between the two $b$-tagged track jets depends on the mass of the underlying resonance, but to first approximation we can assume that the geometry of the fat jet~--- and hence the tagging efficiency~--- depends only on $p_\mathrm{T} / m_J$. In other words, the efficiency to identify a dark Higgs boson with $m_s = \unit[50]{GeV}$ and $p_\mathrm{T} = \unit[400]{GeV}$ will be approximately the same as the efficiency to identify a SM Higgs boson with $p_\mathrm{T} = \unit[1]{TeV}$. We found good agreement with our analysis when using the Higgs-jet tagging efficiencies from~\cite{ATLASperformance} and rescaling them proportional to $m_s / m_h$ in order to estimate the efficiencies for a dark Higgs boson. In this conversion we use loose selection cuts, which apply a less constraining requirement on the fat jet mass.

\subsection{Standard Model backgrounds}

The requirement of two $b$-tagged track jets is sufficient to suppress SM backgrounds from light-flavour fat jets to a negligible level. Even background from fat jets containing $c\bar{c}$ can be reduced by up to a factor of 100~\cite{ATLASperformance}. This leaves two main sources of backgrounds: fat jets containing two $b$-quarks and fat jets containing a $b$-quark and a $c$-quark. The former class of backgrounds results typically from $Z+b\bar{b}$, $W+b\bar{b}$ and diboson events ($ZZ$, $ZW$, and $WW+b\bar{b}$), the latter class is sensitive in particular to pair-produced hadronically decaying top quarks if a $c$-quark is produced in the $W$ boson decay. Even if no $c$-quark is produced, an additional $b$-tagged track jet can arise from QCD radiation in the same fat jet as the original $b$-quark from the top decay, giving a non-negligible contribution to the total background. 

All SM backgrounds are generated with \texttt{MadGraph5\_aMC@NLO~v2.3.3}~\cite{Alwall:2014hca} using the \texttt{NNPDF~v2.3}~\cite{Ball:2012cx} parton distribution functions. For showering, hadronisation and simulation of the underlying event we use \texttt{Pythia~v8.219}~\cite{Sjostrand:2014zea} together with the Monash tune~\cite{Skands:2014pea}. We generate $Z+b\bar{b}$, $ZZ$, $ZW$ at next-to-leading order~(NLO); $W+b\bar{b}$ and $t\bar{t}$ are simulated at leading order~(LO). To include leading effects from QCD radiation for these backgrounds we also generate a sample with an additional jet and merge the two samples as appropriate. Matching uses the MLM scheme with a $k_t$ jet algorithm~\cite{Mangano:2006rw,Alwall:2007fs}. For the NLO samples, showering is done via the \texttt{Pythia~8} interface of \texttt{MadGraph}, LO samples are passed to a stand-alone version of \texttt{Pythia~8}. In the following the different background sources are discussed individually.

$\boldsymbol{Z+b\bar{b}}$\textbf{:} A $b\bar{b}$-pair can recoil against an invisibly decaying $Z$ boson emitted from the initial or final state, leading to an apparent fat jet in association with missing transverse momentum. The distribution of the $b\bar{b}$ invariant mass is continuous but typically peaks at small values. We generate this background at NLO, performing the decay $Z \to \nu \bar{\nu}$ with \texttt{MadSpin}. Since a parton-level $\slashed{E}_\mathrm{T}$ cut cannot be implemented in \texttt{MadGraph} at NLO, we require \mbox{$p_\mathrm{T}(Z) > \unit[300]{GeV}$} instead. 

$\boldsymbol{W+b\bar{b}}$\textbf{:} A dark Higgs decay can also be mimicked by $b\bar{b}$-pairs produced in association with a leptonically decaying $W$ boson, provided the charged lepton is not identified. Including $W+b\bar{b}j$ is essential because this process can take place without a down-type quark (or anti-quark) in the initial state, leading to a significant increase of the total cross section. Therefore we generate (and appropriately merge) $W+b\bar{b}$ and $W+b\bar{b}j$ for quark initial states, as well as $W+b\bar{b}j$ and $W+b\bar{b}jj$ for an initial state consisting of a quark and a gluon. A parton-level requirement of $\slashed{E}_\mathrm{T} > \unit[250]{GeV}$ is applied.

\textbf{Diboson:} Of the various diboson backgrounds, the largest contribution results from $pp \rightarrow Z(\to \text{invisible}) Z(\to b\bar{b})$. However, $pp \rightarrow W(\to \ell\nu) Z(\to b\bar{b})$ where the charged lepton fails identification is also relevant. We generate both $ZZ$ and $ZW$ at NLO and perform the subsequent decays with \texttt{MadSpin}. For the $ZZ$ process we require at least one $Z$ boson with \mbox{$p_\mathrm{T} > \unit[300]{GeV}$}; for $WZ$ we require that the events satisfy \mbox{$p_\mathrm{T}(Z) > \unit[300]{GeV}$} or \mbox{$p_\mathrm{T}(W) > \unit[400]{GeV}$}. Another relevant diboson background is $W^+ W^- +b\bar{b}$ in a gluon-gluon initial state.\footnote{We exclude diagrams from $t \bar{t}$ to avoid double counting.} We simulate this background at LO, using a parton-level cut of \mbox{$\slashed{E}_\mathrm{T} > \unit[400]{GeV}$}.

$\boldsymbol{t \bar{t}}$\textbf{:} Finally, the $t\bar{t}$ background results from semi-leptonic decays of double top production, $t\bar{t} \to b\bar{b} \, W(\to q \bar{q}')W(\to\ell \nu)$, where the charged lepton again fails identification.  At very high $p_\mathrm{T}$, the top-quark is sufficiently boosted for its decay products to be merged into one fat jet. A hadronically decaying top quark may mimic a dark Higgs jet if either the $c$-quark from the $W$ decay is misidentified as a $b$-quark, or if QCD radiation leads to another $b$-tagged track jet. In this case, the fat jet will have an invariant mass close to the top-quark mass, making it easy to distinguish this case from the lighter dark Higgs jets. However, at somewhat smaller $p_\mathrm{T}$ the decay products will result in separate jets and the invariant mass of the fat jet containing the $b$-quark can be different.\footnote{In principle, there could also be a contribution at lower fat jet masses from off-shell top-quarks. This contribution however is found to be negligible. In what follows we therefore only consider the contribution from on-shell top quarks.} NLO event generation in \texttt{MadGraph} does not allow to define a parton level $\slashed{E}_\mathrm{T}$ cut, so a very large number of Monte Carlo~(MC) events is necessary to populate the region of interest of $\slashed{E}_\mathrm{T} > \unit[500]{GeV}$. We therefore generate the $t\bar{t}$ background at LO, applying the requirement $\slashed{E}_\mathrm{T} > \unit[400]{GeV}$ at parton level.

\begin{table}[tbp]
\centering
\begin{tabular}{ccccc}
\hline
\hline
& $t\bar{t}$ & $W+b\bar{b}$ & $Z+b\bar{b}$ & Diboson \\
\hline
Simulation & $2.83 \pm 0.12$ & $1.16 \pm 0.06$ & $2.42 \pm 0.07$ & $0.56 \pm 0.02$ \\
ATLAS prediction& $4.83 \pm 0.88$ & $2.48 \pm 0.71$ & $3.80 \pm 0.44$ & $1.20 \pm 0.12$ \\
\hline
Rescaling factor & $1.7 \pm 0.3$ & $2.1 \pm 0.6$ & $1.6 \pm 0.2$ & $2.1 \pm 0.2$ \\
\hline
\hline
\end{tabular}
\caption{\label{tab:rescaling} Predicted number of events with $\unit[80]{GeV} \leq m_J \leq \unit[280]{GeV}$ according to our simulation and in ATLAS~\cite{ATLASanalysis}, as well as the resulting rescaling factors. For the results of our simulation we quote the uncertainties resulting from limited MC statistics. The uncertainties in the rescaling factors reflect the uncertainties in the background estimates from~\cite{ATLASanalysis}.} 
\end{table}

To validate our background simulation we consider the same search region as discussed in~\cite{ATLASanalysis}, i.e.\ $\slashed{E}_\mathrm{T} > \unit[500]{GeV}$ and $\unit[80]{GeV} \leq m_J \leq \unit[280]{GeV}$. While we observe good shape agreement of the various backgrounds, the number of predicted events in the signal region is underestimated by a factor 1.5--2 for the various contributions (see table~\ref{tab:rescaling}). This level of discrepancy is not unexpected given that for some of the backgrounds we estimate higher-order corrections by including a leading QCD jet and use a relatively simple detector simulation and fat jet tagging efficiency. The largest discrepancy is found for the $W+bb$ background, which is difficult to simulate and where large $K$-factors are expected~\cite{Cordero:2009kv,Luisoni:2015mpa}. The reason is that the dominant contribution arises from $qg\rightarrow W+b\bar{b}q'$, which is of higher order in $\alpha_s$ and therefore very sensitive to scale uncertainties.

\begin{figure}[t]
  \centering
    \includegraphics[width=0.45\textwidth]{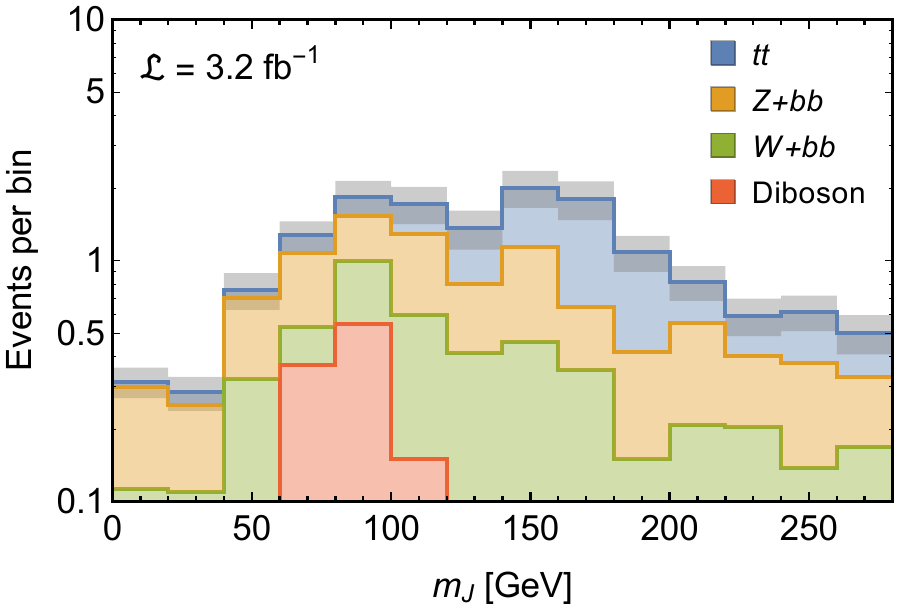}\qquad
    \includegraphics[width=0.45\textwidth]{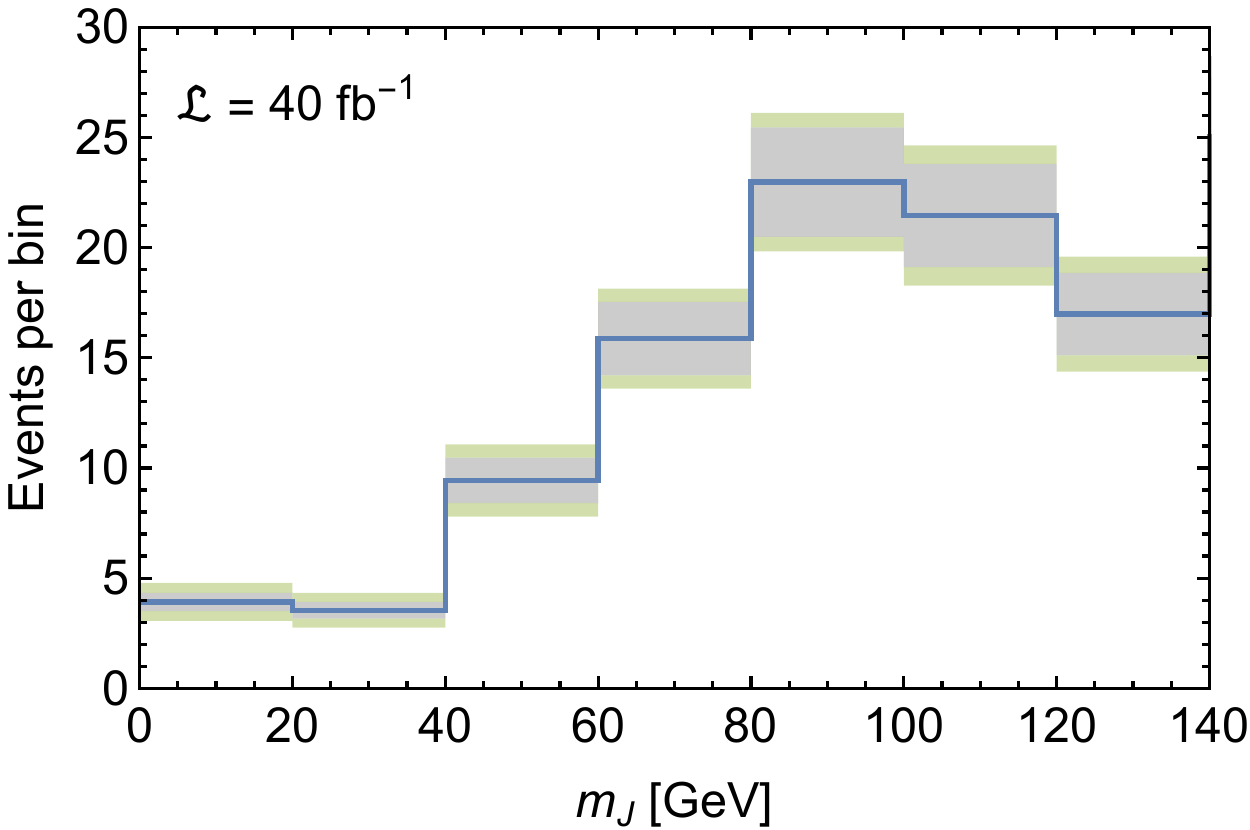}
  \caption{Left: predicted number of events per bin for different sources of background as a function of $m_J$ for an integrated luminosity of $\unit[3.2]{fb^{-1}}$, to be compared with~\cite{ATLASanalysis}. Right: predicted total number of background events per bin as a function of $m_J$ in the low $m_J$ search window for an integrated luminosity of $\unit[40]{fb^{-1}}$. In both cases, events with $\slashed{E}_\mathrm{T} < \unit[500]{GeV}$ are rejected. The gray band in both panels indicates the uncertainties resulting from the rescaling factors, the green band in the right panel includes in addition the uncertainties resulting from limited MC statistics.}
\label{fig:background}
\end{figure}

We apply scale factors to each background to reproduce the predicted yields in \cite{ATLASanalysis}. The quoted uncertainties of the background estimates can then be propagated to uncertainties on the rescaling factor. The results are summarised in table~\ref{tab:rescaling}. When summing up the rescaled background contributions the shape of the total predicted background shown in the left panel of figure~\ref{fig:background} is in excellent agreement with the ATLAS analysis.

We use the same rescaling factors to estimate the backgrounds at lower values of $m_J$. We find that the $t\bar{t}$ background decreases very rapidly in this regime and the dominant background contributions come from $Z+b\bar{b}$. To illustrate sensitivities achievable we consider an integrated luminosity of $\unit[40]{fb^{-1}}$. We estimate the expected background uncertainties by assuming that the current background estimates have a systematical uncertainty of $10\%$ that will remain unchanged for $\unit[40]{fb^{-1}}$. The remaining uncertainty is assumed to result from limited statistics and is therefore expected to scale with $1/\sqrt{\mathcal{L}}$ with increasing luminosity $\mathcal{L}$. The resulting background prediction is shown in the right panel of figure~\ref{fig:background}. In the region of interest we predict between 10 and 25 background events per bin of \unit[20]{GeV}, with an uncertainty of approximately 11\%. This number increases to about 15\% when including the uncertainties from MC statistics. However in either case the uncertainties of the background prediction are significantly smaller than the expected Poisson fluctuations. Hence the experimental sensitivity is still statistically limited.

\section{Signal prediction and expected sensitivity}
\label{sec:sensitivity}

For the purpose of signal generation the interaction Lagrangians from eqs.~(\ref{eq:Ldark}) and (\ref{eq:Lquark}) are implemented in \texttt{FeynRules~v2.0}~\cite{Alloul:2013bka} in order to produce a \texttt{UFO} model file~\cite{Degrande:2011ua}. We generate events containing a pair of DM particles and a dark Higgs boson at leading order using \texttt{MadGraph5\_aMC@NLO~v2.3.3}~\cite{Alwall:2014hca}, allowing for one additional jet from initial state radiation. To avoid  generation of events with low missing transverse momentum we impose a parton-level cut of $\slashed{E}_\mathrm{T} > \unit[400]{GeV}$, well below the analysis selection. We use \texttt{Pythia 8.219}~\cite{Sjostrand:2014zea} for showering and apply the same \texttt{Rivet} analysis as for the background estimation described in section~\ref{sec:backgrounds}.

\begin{figure}[tb]
\centering
\includegraphics[width=0.45\textwidth]{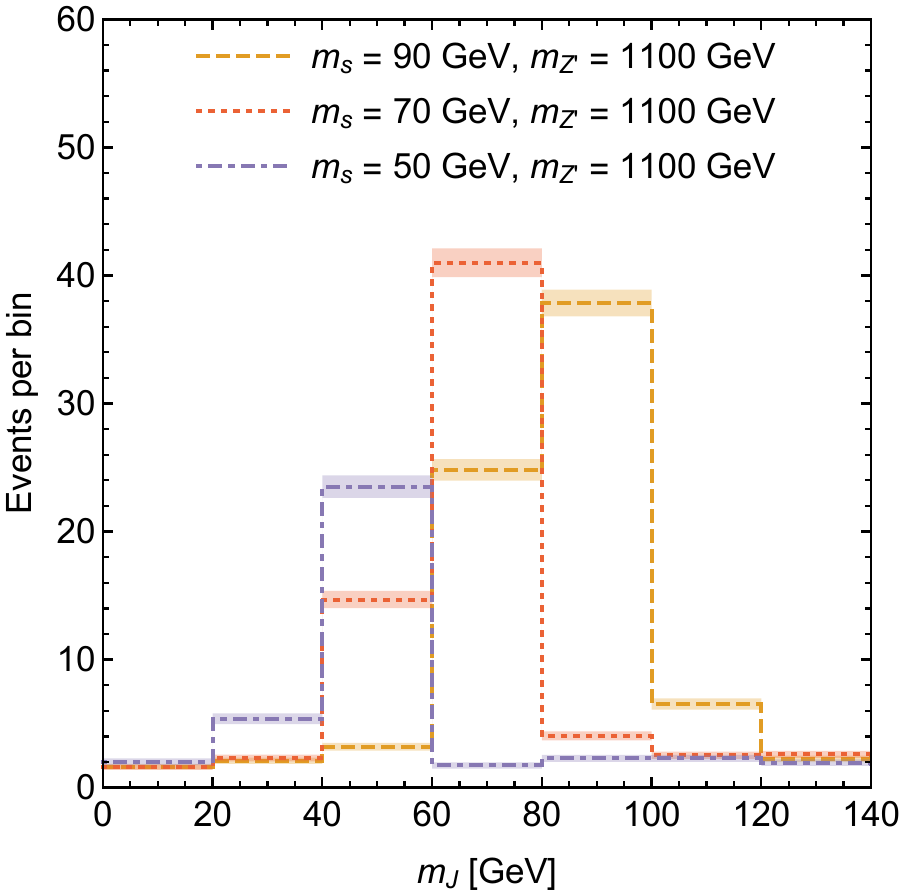}
\includegraphics[width=0.45\textwidth]{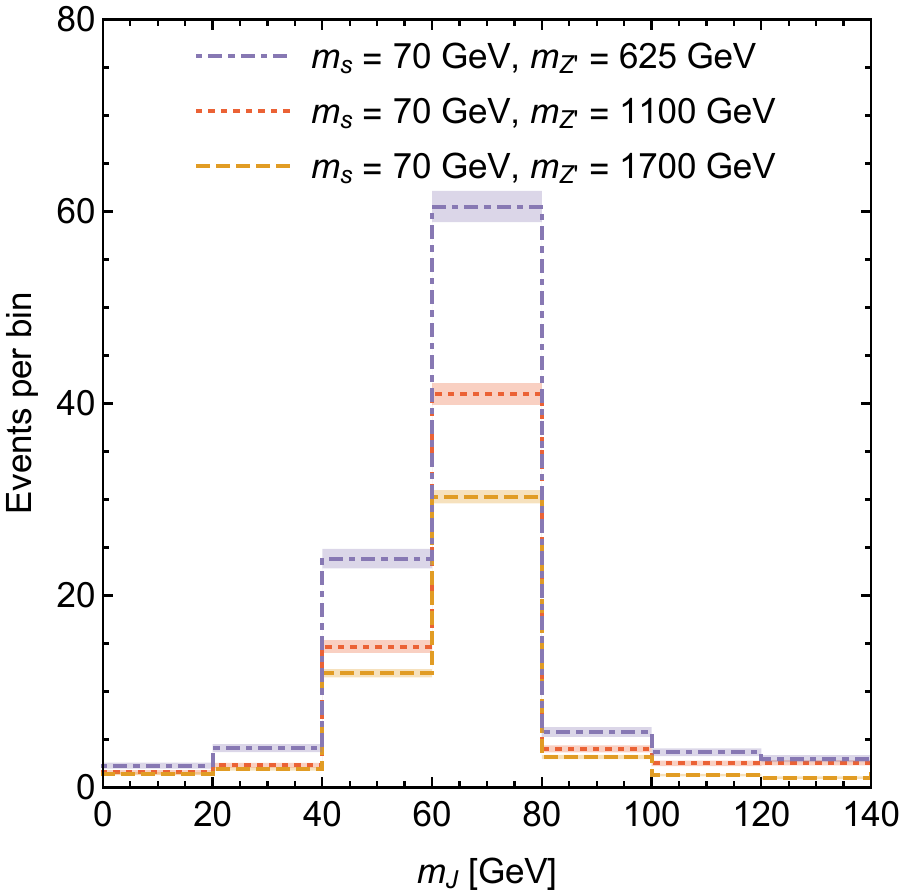}
\caption{Distribution of $m_J$ for signal models with different dark Higgs mass (left) and with different $Z'$ mass (right) and an integrated luminosity of $40\:\text{fb}^{-1}$. In both figures we only consider events with $\slashed{E}_\mathrm{T} > \unit[500]{GeV}$.}
\label{fig:signal}
\end{figure} 

Figure~\ref{fig:signal} shows the fat jet mass distribution for a number of different signal hypotheses, again imposing a cut on the missing transverse momentum of $\slashed{E}_\mathrm{T} > \unit[500]{GeV}$. This ensures that the dark Higgs boson is produced with large transverse momentum and its decay products are boosted into a single fat jet. Therefore the invariant mass of the leading fat jet exhibits a clear peak close to the mass of the dark Higgs boson. Although the width of this peak is relatively wide, there is still a striking difference in the shape of the signal and background distributions.

\begin{table}
\begin{center}
\begin{tabular}{lccc}
\toprule
 & $p_\mathrm{T}(j_1) > \unit[250]{GeV}$ & Dark Higgs tagged & $\unit[40]{GeV} \leq m_J \leq \unit[80]{GeV}$ \\
\midrule
Background & $14063 \pm 790$ & $193 \pm 21$ & $25.3 \pm 3.4$ \\
$m_{Z'} = \unit[0.5]{TeV}$ & 5015 + 363 & 124 & 88.6 \\
$m_{Z'} = \unit[1]{TeV}$ & 1448 + 274 & 88.4 & 60.5 \\
$m_{Z'} = \unit[2]{TeV}$ & 158 + 116 & 39.1 & 27.6 \\
\bottomrule
\end{tabular}
\end{center}
\caption{Cut flow for signal and background events. The entries give the expected number of events with $\slashed{E}_\mathrm{T} > \unit[500]{GeV}$ at $\unit[40]{fb^{-1}}$ for successively more stringent cuts on the leading jet. For the background predictions we also provide the combined statistical and systematical uncertainties. For all signal predictions, we assume $m_\chi = \unit[100]{GeV}$, $m_s = \unit[70]{GeV}$, $g_q = 0.25$ and $g_\chi = 1$. For the signal predictions, the numbers in the first column give the mono-jet $+$ mono-dark-Higgs event numbers. The number of mono-jet events passing the dark Higgs tagging requirement is negligible.}
\label{tab:cutflow}
\end{table}

Table~\ref{tab:cutflow} illustrates the impact of the dark Higgs tagging and the importance of exploiting the shape of the $m_J$ distribution. The first column lists the expected number of background and signal events for an integrated luminosity of $40\:\text{fb}^{-1}$ in the case that no jet tagging is applied, i.e.\ considering all events with a high-$p_\mathrm{T}$ jet, $\slashed{E}_\mathrm{T} > \unit[500]{GeV}$ and the mono-jet selection discussed in section~\ref{sec:framework}.\footnote{For the background expectation this number is obtained by rescaling the background estimate from the most recent ATLAS mono-jet search~\cite{Aaboud:2016tnv}. For the signal prediction, we quote separately those events that do not contain a dark Higgs boson (i.e.\ conventional mono-jet events) and those events that do.} The second column corresponds to the number of background and signal events that remain after applying the jet tagging described in section~\ref{sec:backgrounds}. Finally, the third column indicates the number of events in which the fat jet mass falls within a window around the dark Higgs mass assumed for the event generation (here $\unit[70]{GeV}$). These requirements are found to remove more than $99.8\%$ of all background events with large missing transverse momentum with a signal efficiency of up to $20\%$. After full event selection including the cut in $m_J$ the signal-to-background ratio is 3.5, 2.4 and 1.1 for $Z'$ masses of 0.5, 1 and $\unit[2]{TeV}$, respectively. All three signal hypotheses considered in table~\ref{tab:cutflow} should therefore be easily distinguishable from the dominant backgrounds. 

To calculate the expected sensitivity of possible LHC searches we exploit the shape of the $m_J$ distribution to discriminate signal from background, i.e.\ we omit the $m_J$ window cut from table~\ref{tab:cutflow}. We use \texttt{RooStats} \cite{Moneta:2010pm} to implement the $CL_s$ method~\cite{Read:2002hq} in the asymptotic limit~\cite{Cowan:2010js}, introducing a separate nuisance parameter for the normalisation of each background contribution. This enables us to calculate the expected $p$-value for a given signal hypothesis and determine all parameter points that can be excluded at the 95\% confidence level.

For comparison we also calculate the expected sensitivity of a traditional search for high-$p_\mathrm{T}$ jets in association with missing transverse momentum, where no additional requirements on the leading jet are imposed. For this purpose, we make use of a \texttt{Madanalysis5}~\cite{Conte:2012fm,Conte:2014zja} implementation~\cite{monojet} of the most recent ATLAS mono-jet search~\cite{Aaboud:2016tnv}. We scale the existing experimental results to $\unit[40]{fb^{-1}}$ under the assumption that relative systematic uncertainties remain the same, while relative statistical errors decrease with~$1/\sqrt{\mathcal{L}}$ with increasing luminosity $\mathcal{L}$. Signal events are generated using \texttt{MadGraph5\_aMC@NLO~v2.4.3} again applying a parton level cut of $\slashed{E}_\mathrm{T} > \unit[400]{GeV}$. Detector effects are simulated with \texttt{Delphes~v3}~\cite{deFavereau:2013fsa}. We note that, since our study is based on a Majorana DM candidate rather than a Dirac DM candidate, the expected sensitivity is somewhat lower than the one found in the most recent CMS mono-jet search based on $\unit[12.9]{fb^{-1}}$~\cite{CMS:2016pod}.

\begin{figure}[tb]
\centering
\includegraphics[width=0.46\textwidth]{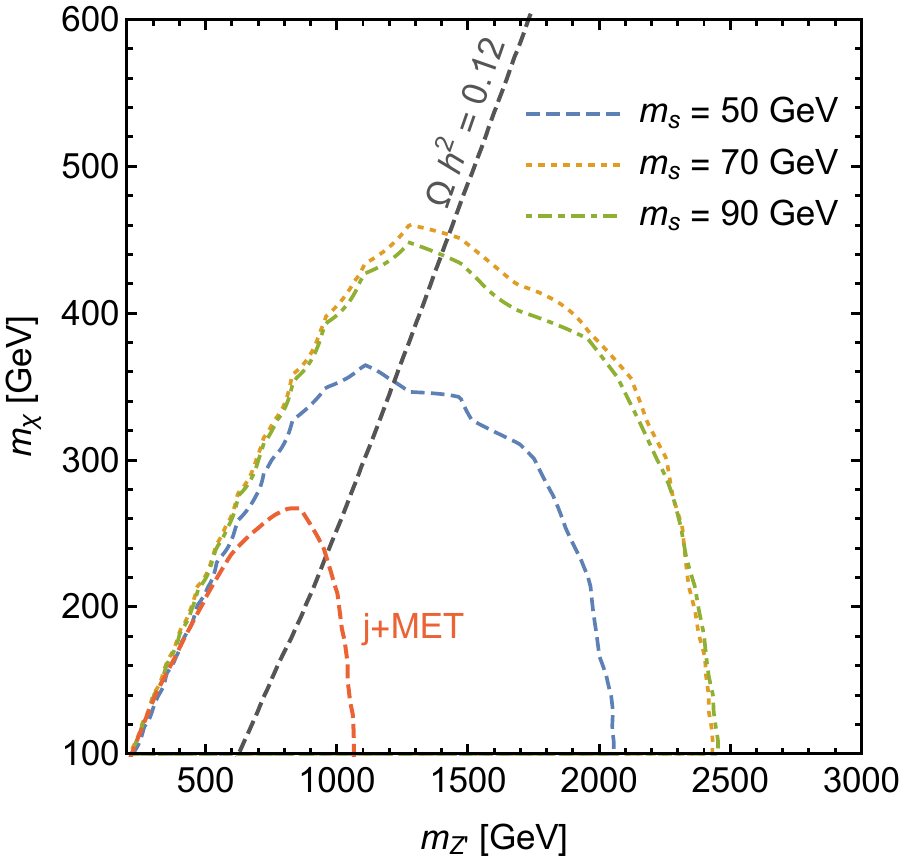}\qquad
\includegraphics[width=0.45\textwidth]{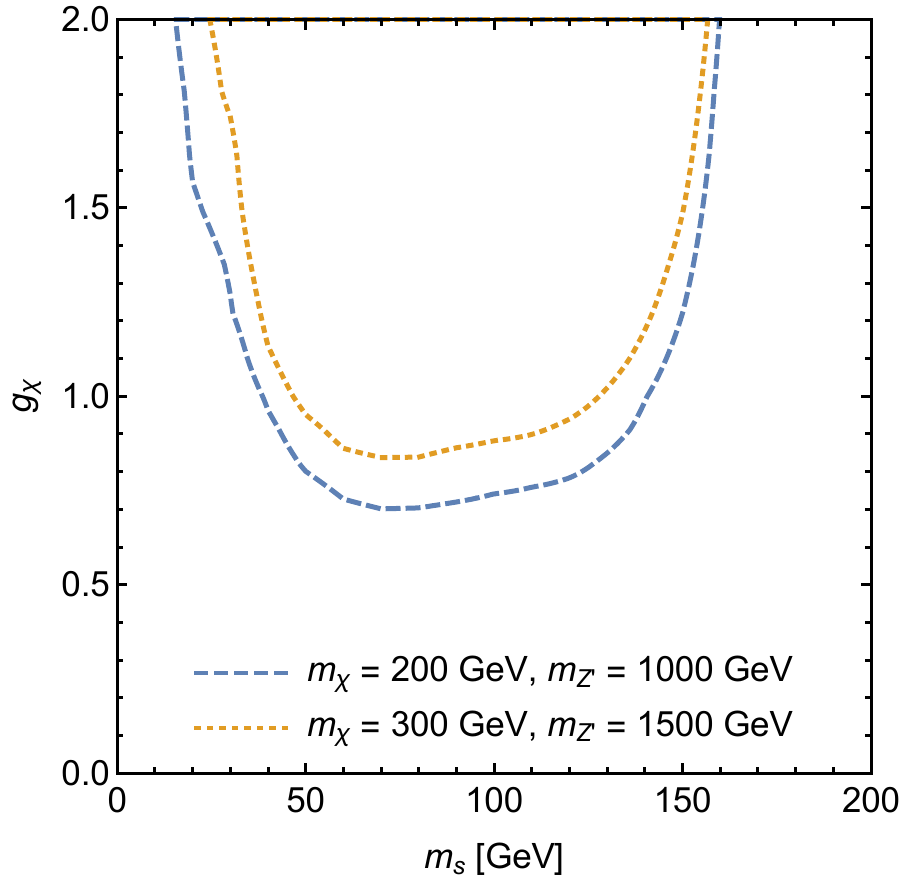}
\caption{Left: expected sensitivity of a mono-dark-Higgs search with an integrated luminosity of $\unit[40]{fb^{-1}}$, considering events with $\slashed{E}_\mathrm{T} > \unit[500]{GeV}$ and using a $CL_s$ method to compare the $m_J$-distribution for signal and background. Dashed, dotted and dash-dotted lines correspond to $m_s = \unit[50]{GeV}$, $m_s = \unit[70]{GeV}$ and $m_s = \unit[90]{GeV}$, respectively. For comparison, we show the expected sensitivity of a conventional mono-jet search and the parameter combinations for which the observed relic abundance is reproduced. Right: bound on the dark sector coupling $g_\chi$ as a function of the dark Higgs mass $m_s$ for two different benchmark scenarios.}
\label{fig:sensitivity}
\end{figure}

Our results are shown in the left panel of figure~\ref{fig:sensitivity} for our choice of couplings ($g_q = 0.25$, $g_\chi = 1$) and for different values of the dark Higgs mass. We also show the combinations of masses that reproduce the observed relic abundance, $\Omega h^2 \approx 0.12$~\cite{Ade:2015xua}.\footnote{The exact position of this line depends slightly on the dark Higgs mass $m_s$. For definiteness we take $m_s= \unit[70]{GeV}$.} We find that the dedicated search for a mono-dark-Higgs signal can probe ranges of the $Z'$ mass and the DM mass that are inaccessible to conventional mono-jet searches. For $m_s = \unit[70]{GeV}$ and $m_s = \unit[90]{GeV}$ the expected sensitivity is almost identical, extending up to $m_{Z'} = \unit[2500]{GeV}$ and $m_\chi = \unit[450]{GeV}$. 

For $m_s = \unit[50]{GeV}$ we find a somewhat lower expected sensitivity. The reason is that in this case the boost factor of the dark Higgs boson becomes so large that the two $b$-jets from its decay merge into a single track jet and the dark Higgs boson can no longer be correctly identified. Indeed, the dark Higgs tagging efficiency~--- and therefore the sensitivity~--- drops rapidly for dark Higgs masses below $\unit[50]{GeV}$. This observation is illustrated in the right panel of figure~\ref{fig:sensitivity}, where we show the upper bound on the dark sector coupling $g_\chi$ as a function of the dark Higgs mass $m_s$ for two relevant benchmark scenarios. 

The right panel of figure~\ref{fig:sensitivity} also shows that the mono-dark-Higgs search developed above can equally be used to search for dark Higgs bosons with masses above the mass of the SM Higgs boson, provided the $Z'$ is sufficiently heavy to allow the decay $Z' \rightarrow \chi \chi s$ and that sufficient $\slashed{E}_\mathrm{T}$ can be produced. The dark Higgs tagging efficiency is essentially constant for $\unit[50]{GeV} < m_s < \unit[150]{GeV}$ and the slight weakening of the bound for increasing $m_s$ reflects the reduced probability of dark-Higgs strahlung. For $m_s \gtrsim \unit[160]{GeV}$ the decay mode $s \to W^+W^-$ becomes accessible and decays into bottom quarks become strongly suppressed.

To conclude this section, we note that it may be possible to improve the sensitivity to very light dark Higgs bosons by loosening the cut on the missing transverse momentum so that dark Higgs bosons with lower boost factors can contribute. Conversely, for heavier dark Higgs bosons and large $Z'$ masses it is conceivable that a harder $\slashed{E}_\mathrm{T}$-cut can further enhance the sensitivity. A detailed study of the optimal missing transverse momentum cut as a function of the masses of the various particles is beyond the scope of this work.

\section{Dark matter relic density}
\label{sec:relic}

So far we have limited ourselves to a specific choice of couplings, namely $g_q = 0.25$ and $g_\chi = 1$. While this allows to make contact with existing DM searches at the LHC, it has a number of important drawbacks. First, fixing the couplings to specific values means that the observed DM relic abundance is only reproduced for certain combinations of the masses of the particles in the dark sector. Second, the specific coupling combination adopted so far is in fact excluded by searches for di-jet resonances for a wide range of $Z'$ masses~\cite{Chala:2015ama,Fairbairn:2016iuf,ATLAS:2016bvn,ATLAS:2016lvi,Sirunyan:2016iap}. In this section we discuss how requiring to reproduce the thermal DM relic abundance changes the sensitivities of various searches in a non-trivial way, diminishing the importance of di-jet searches while at the same time enhancing the sensitivity of dark Higgs searches.

\begin{figure}[tb]
\centering
\includegraphics[width=0.45\textwidth]{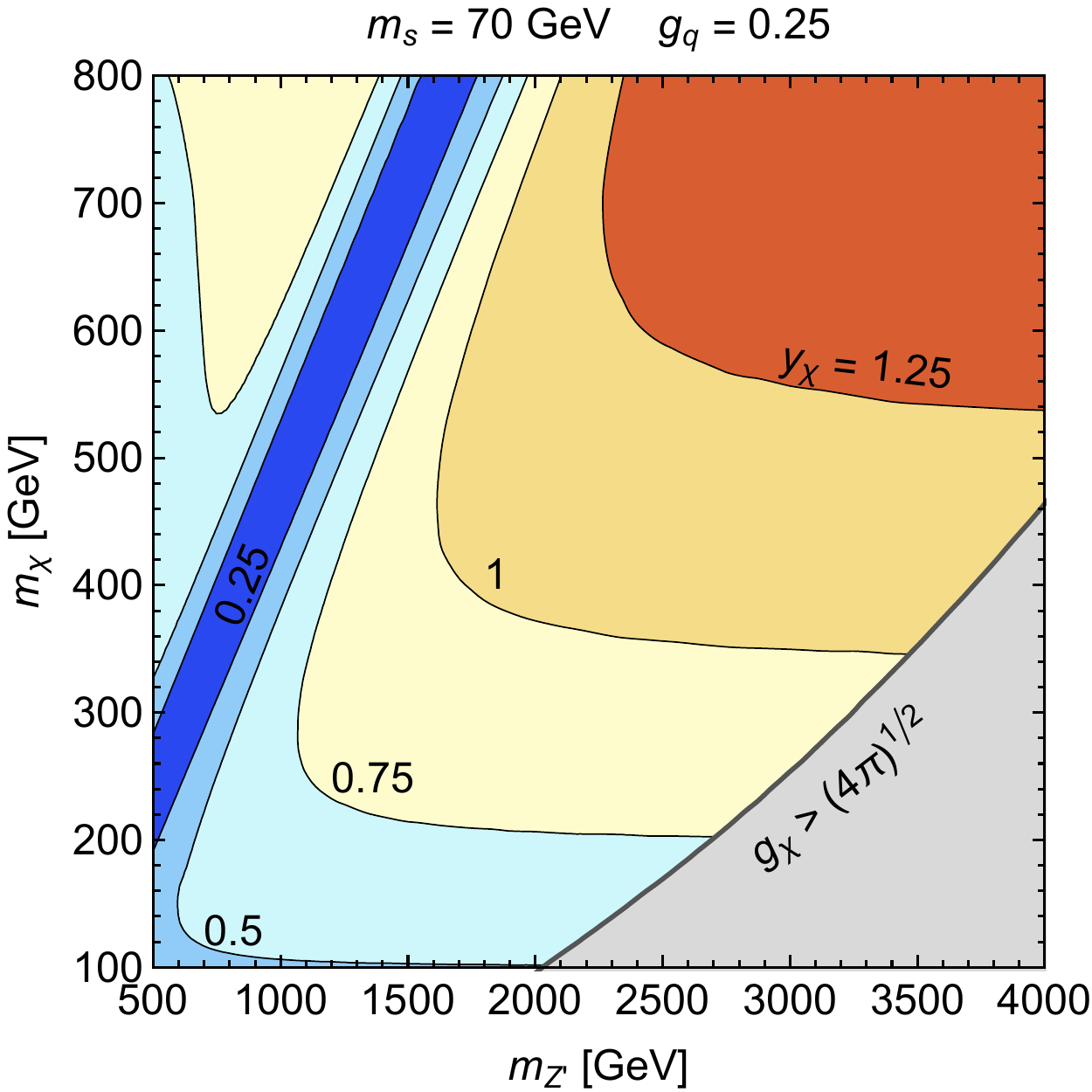}\qquad\includegraphics[width=0.45\textwidth]{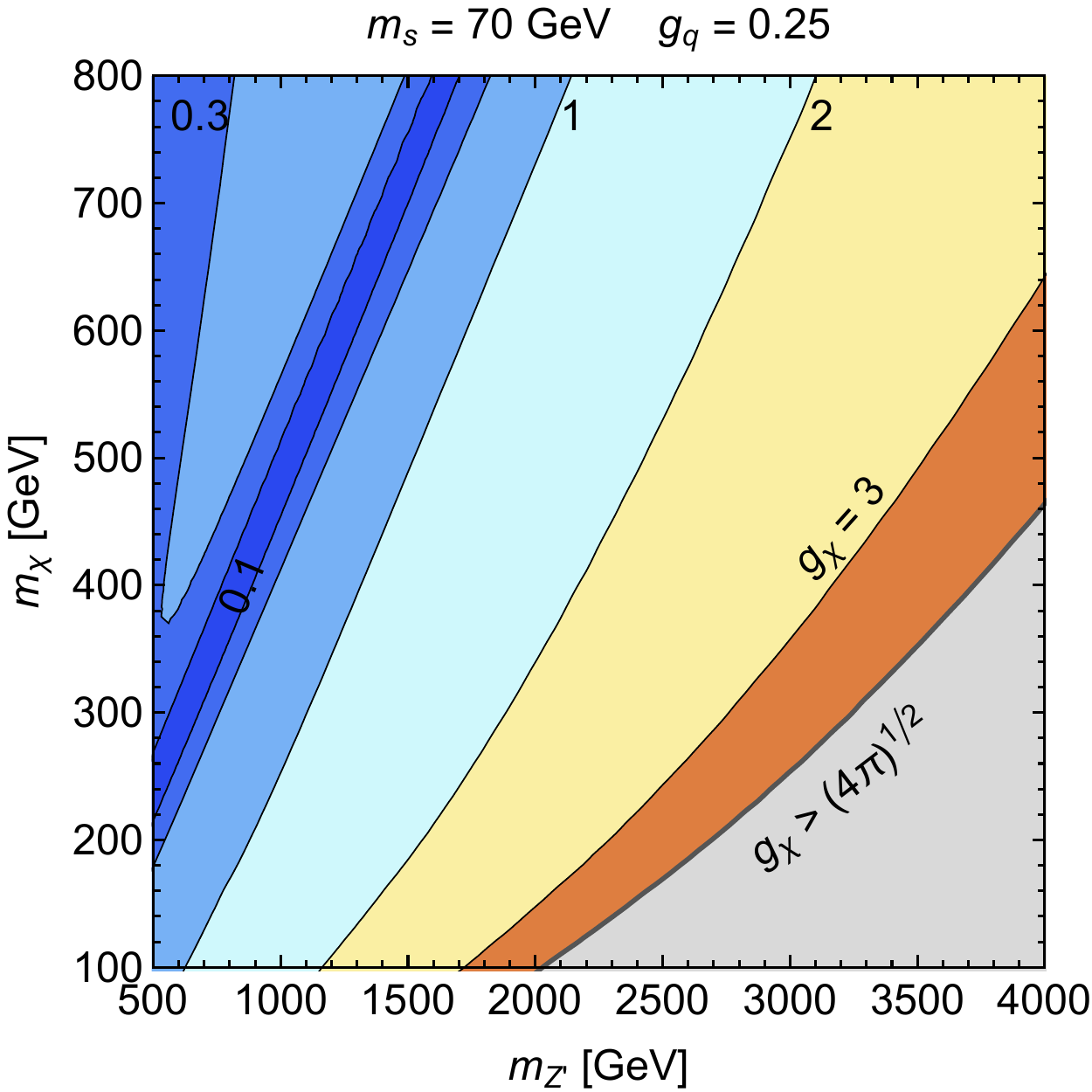}
\caption{Value of $y_\chi$ (left) and $g_\chi$ (right) in the $m_\chi$ -- $m_{Z'}$ plane determined by the relic abundance requirement for $m_s = \unit[70]{GeV}$ and $g_q = 0.25$.}
\label{fig:gxrelic}
\end{figure} 

The amount of DM in the Universe is well known,  $\Omega h^2 \approx 0.12$~\cite{Ade:2015xua}, and any model of DM should be consistent with this observation.
In particular DM should not be overproduced, which is a challenge for models with only one heavy mediator. 
An attractive feature of the model we study is that the DM relic abundance can be explained via thermal freeze-out in the early Universe in most of the parameter space.
The crucial observation is that for $m_s < m_\chi$ the process $\chi \chi \to s s$ followed by $s \to \text{SM}$ provides a possible DM annihilation channel. 
While the process $\chi \chi \rightarrow Z' \rightarrow q\bar{q}$ becomes inefficient for $m_{Z'} \gg m_\chi$, annihilation into dark Higgs bosons allows to avoid DM overproduction also in this parameter region.

All DM annihilation processes depend either on the DM Yukawa coupling $y_\chi$ (if they involve dark Higgs bosons) or on the DM gauge coupling $g_\chi$ (if they involve $Z'$ bosons). As discussed in section~\ref{sec:framework}, if the dark Higgs is responsible for generating both the DM mass and the $Z'$ mass, these two couplings are in fact related by
\begin{equation}
g_\chi = \frac{m_{Z'}}{m_\chi} \frac{y_\chi}{2 \sqrt 2} \; .
\label{eq:gxyx}
\end{equation}
Requiring the predicted relic abundance from thermal freeze-out to agree with observations therefore unambiguously determines both couplings for given masses and SM couplings. Figure~\ref{fig:gxrelic} shows the values of $y_\chi$ and $g_\chi$ obtained this way for $m_s = \unit[70]{GeV}$ and $g_q = 0.25$. To calculate the relic density we use \texttt{micrOMEGAs~v4.2.5}~\cite{Belanger:2014vza}.

For $m_\chi \approx m_{Z'} / 2$ the process $\chi \chi \rightarrow Z' \rightarrow q\bar{q}$ is resonantly enhanced and therefore yields the dominant contribution to the DM annihilation cross section. Very small values of $g_\chi$ are then sufficient to reproduce the observed relic abundance and the process $\chi \chi \to s s$ is not relevant. Nevertheless in the non-resonant regime and in particular for large $Z'$ masses direct annihilation into SM quarks becomes inefficient and $\chi \chi \rightarrow s s \rightarrow \text{SM}$ becomes the leading annihilation process. 
Accordingly the inferred value of $y_\chi$ becomes independent of $m_{Z'}$ for $m_{Z'} \gg m_\chi$ (see left panel of figure~\ref{fig:gxrelic}). For example, for $m_\chi = \unit[200]{GeV}$, $m_s = \unit[70]{GeV}$ and $m_{Z'} \gg m_\chi$ the DM relic abundance is reproduced for $y_\chi \approx 0.75$. It follows from eq.~(\ref{eq:gxyx}) that for fixed $y_\chi$ and fixed $m_\chi$, larger $Z'$ masses imply larger values of $g_\chi$ (see right panel of figure~\ref{fig:gxrelic}). In the example above, one obtains $g_\chi \approx 2$ ($g_\chi \approx 3.3$) for $m_{Z'} = \unit[1.5]{TeV}$ ($m_{Z'} = \unit[2.5]{TeV}$).

This observation leads to three important consequences. First, we conclude that the parameter space is bounded from above by the requirement that the $Z'$--DM coupling remains perturbative~\cite{Chala:2015ama,Kahlhoefer:2015bea,Duerr:2016tmh}. For $m_\chi = \unit[200]{GeV}$ and \mbox{$m_s = \unit[70]{GeV}$}, for example, the requirement $g_\chi < \sqrt{4\pi}$ implies the bound $m_{Z'} \lesssim \unit[2.7]{TeV}$. 
The second consequence is that the probability for a $Z'$ to emit a dark Higgs boson is enhanced for heavy $Z'$ bosons. This is because the coupling between the $Z'$ and the dark Higgs boson is proportional to $g_\chi$ and therefore, for constant $y_\chi$, proportional to $m_{Z'}$. If we determine the dark sector couplings via the relic density requirement we expect significantly larger signal rates compared to the benchmark case considered in 
section~\ref{sec:sensitivity}. Finally, a larger $Z'$--DM coupling also implies a larger partial decay width for the process $Z' \rightarrow \chi \chi$. As a result the invisible branching ratio of the $Z'$ increases and decays to quarks are suppressed. This effect reduces di-jet signal rates and can therefore potentially hide the model from direct searches for the $Z'$ mediator. The larger coupling also leads to a broader resonance which makes it even more difficult to distinguish a potential di-jet signal from QCD backgrounds.

\begin{figure}[tb]
\centering
\includegraphics[width=0.45\textwidth]{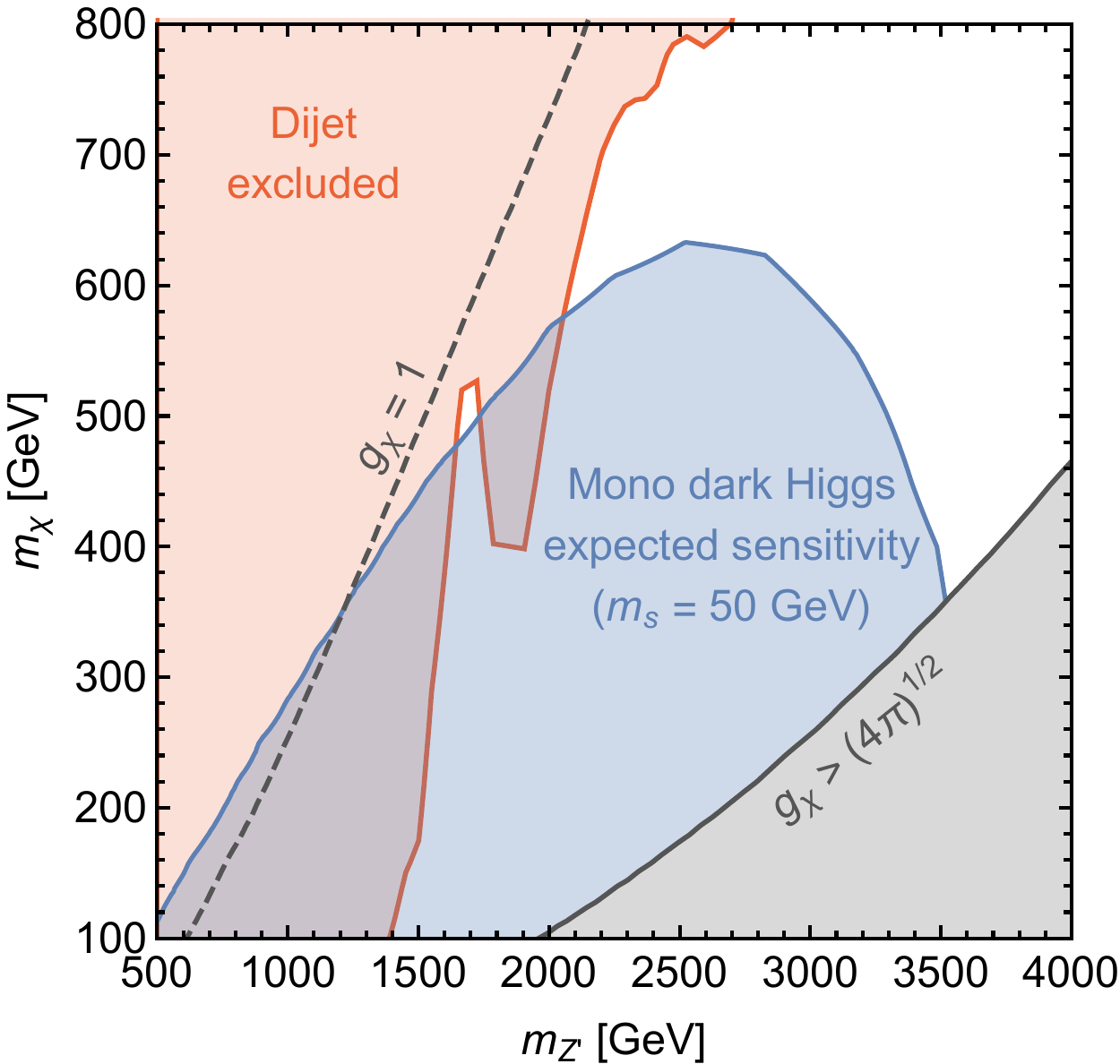}\qquad\includegraphics[width=0.45\textwidth]{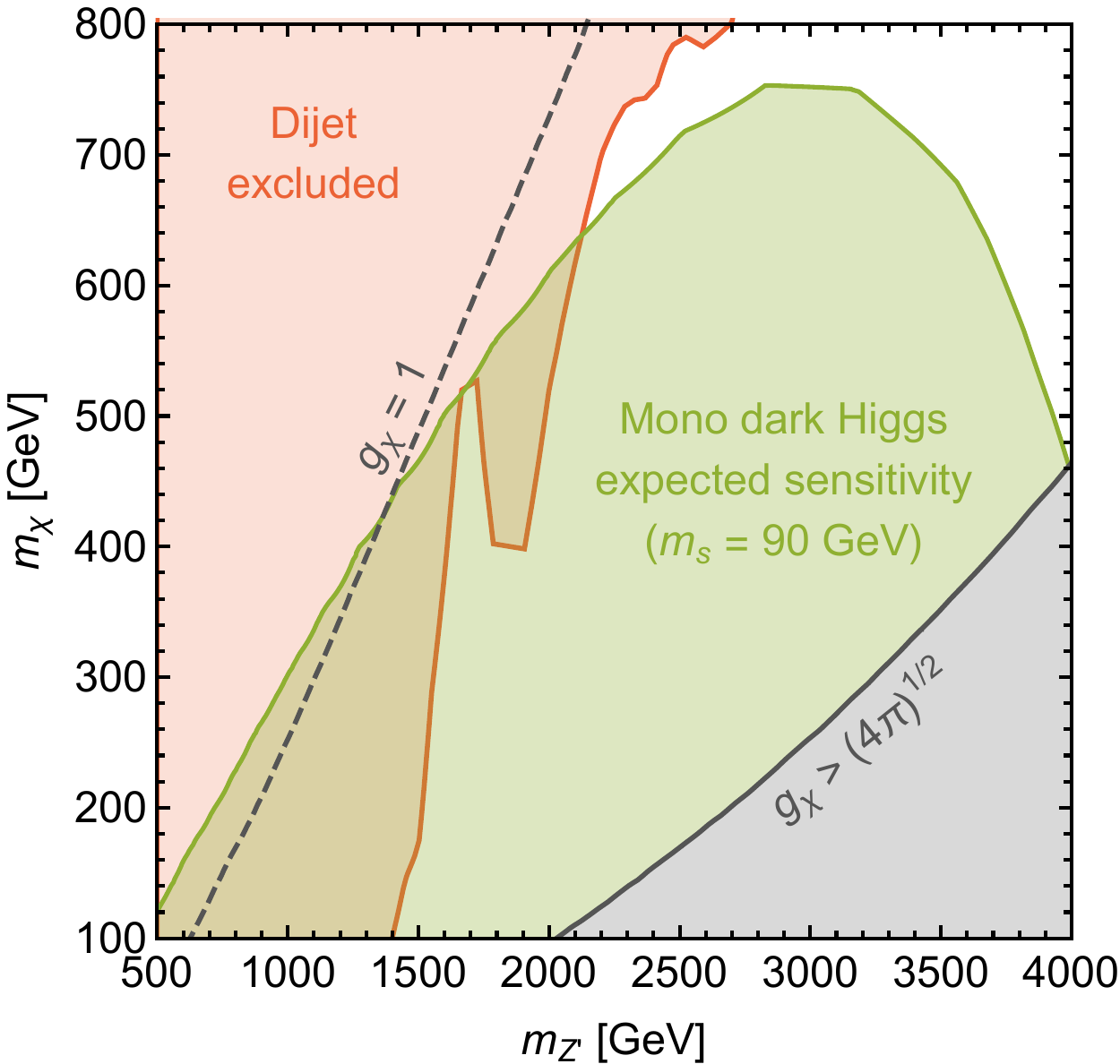}
\caption{Expected sensitivity of a search for dark Higgs bosons with $m_s = \unit[50]{GeV}$ (left) and \mbox{$m_s = \unit[90]{GeV}$} (right) in the $m_{Z'}$--$m_\chi$ parameter plane for $g_q = 0.25$ and under the assumption that the dark sector coupling $y_\chi$ (or equivalently $g_\chi$) is fixed by the relic density requirement (see figure~\ref{fig:gxrelic}). For comparison, we show existing bounds from a search for di-jet resonances that are obtained from~\cite{Fairbairn:2016iuf}. By construction, the line labelled $g_\chi = 1$ is identical to the one labelled $\Omega h^2 = 0.12$ in the left panel of figure~\ref{fig:sensitivity}.}
\label{fig:relicbound}
\end{figure} 

In other words, if we fix the DM Yukawa coupling in such a way that DM overproduction is avoided in the parameter region with $m_{Z'} \gg m_\chi$, the model automatically predicts an enhancement of mono-dark-Higgs signals and a suppression of di-jet signals.\footnote{We note that allowing for an underabundance, i.e.\ assuming that $\chi$ constitutes only part of the DM, the values of $g_\chi$ and $y_\chi$ inferred above would correspond to a lower bound. With such an assumption the expected dark Higgs signal would therefore be even larger.} This is illustrated in figure~\ref{fig:relicbound}, where we show the expected sensitivity of the mono-dark-Higgs search introduced above for the case that the DM Yukawa coupling is determined from the relic density. We observe that for sufficiently small DM masses this search is sensitive to $Z'$ masses up to the perturbativity bound~\cite{Chala:2015ama,Kahlhoefer:2015bea}. For comparison we also show existing bounds from di-jet resonance searches, obtained from a combination of several different di-jet searches at ATLAS and CMS~\cite{Fairbairn:2016iuf}. As expected, these bounds are suppressed in the region with $g_\chi > 1$, where the mono-dark-Higgs search is most sensitive, leading to an appealing complementarity of the two search strategies.

\section{Conclusion}
\label{sec:conclusions}

We have presented expected sensitivities of LHC searches for additional light Higgs bosons produced in association with DM. Such searches are well-motivated for two reasons: First, a dark Higgs boson is a natural component of models of DM coupled to a spin-1 mediator. It allows to generate the masses of the particles in the dark sector in a gauge-invariant way and is in fact necessary to restore unitarity at high energies. Second, the case that the dark Higgs boson is the lightest particle in the dark sector offers an attractive way to set the DM relic abundance via the process $\chi \chi \to s s$ followed by decays of the dark Higgs boson into SM states. This also naturally avoids the DM overproduction predicted by $Z'$ exchange alone.

If both the DM particle and the dark Higgs boson are light compared to the $Z'$ boson, the three-body decay $Z' \to \chi \chi s$ can lead to a highly-energetic dark Higgs boson. The dark Higgs boson is expected to decay preferentially to a pair of boosted bottom quarks. The resulting experimental signature is hence a single fat jet containing two $b$-tagged subjets in association with large missing transverse momentum.

By employing refined jet tagging techniques experimental backgrounds can be substantially reduced. We have presented an estimate of the expected backgrounds based on MC simulations and a comparison with existing experimental studies. Based on these estimates we then derive sensitivities for specific signal models.

For the benchmark case of a $Z'$ coupling to DM with $g_\chi = 1$ and to quarks with $g_q = 0.25$, we find that with already collected data LHC searches can be sensitive to $Z'$ masses up to  $\unit[2500]{GeV}$ and DM masses up to $\unit[500]{GeV}$ for dark Higgs masses in the range $\unit[50]{GeV} \lesssim m_s \lesssim \unit[150]{GeV}$. For $m_s > 2 m_W$ the search strategy loses sensitivity because decays into $W^+W^-$ suppress the branching ratio for the dark Higgs boson to decay into bottom quarks. Searches for diboson resonances in association with missing transverse momentum may provide an interesting opportunity to explore also this parameter region.

For dark Higgs masses below $\unit[50]{GeV}$ the experimental sensitivity is reduced because the boost is so large that the two $b$-jets become indistinguishable from each other. Higher sensitivity might be achieved by lowering the $\slashed{E}_\mathrm{T}$-cut at the expense of increasing backgrounds. We note, however, that very small dark Higgs masses are independently constrained by other searches: as soon as $m_s < m_h/2$, the SM Higgs boson can decay into two dark Higgs bosons~\cite{Duerr:2016tmh}. While such decays may be difficult to observe directly, they nevertheless may modify the branching ratios of the SM Higgs boson, leading to strong bounds on the mixing angle.

Finally we discussed the case where the dark sector coupling is set by the relic density requirement. We find that this typically requires values of $g_\chi$ larger than unity for $g_q = 0.25$. Such large values of $g_\chi$ imply that the rate of dark-Higgs strahlung is dramatically increased, whereas the sensitivity of di-jet searches is strongly suppressed. In this set-up mono-dark-Higgs searches can potentially probe $Z'$ masses up to perturbative unitarity bounds, and DM masses up to $\unit[800]{GeV}$, whereas searches for di-jet resonances are mostly sensitive to the parameter region where $g_\chi < 1$.

While the precise value of the mixing angle between the SM Higgs boson and the dark Higgs boson is not relevant for this analysis, we have assumed throughout this work that the mixing is sufficiently large for the dark Higgs boson to decay promptly. For even smaller mixing angles the dark Higgs decay can result in a displaced vertex. This promising experimental signature is left for future work.

In conclusion, there are ample theoretical reasons to expect the presence of an additional Higgs boson in the dark sector. If such a dark Higgs decays dominantly into SM states, it may provide us with a unique window to explore the dark sector. The only requirement is the production of any dark sector state with sufficiently large momentum, so that dark-Higgs strahlung becomes sizeable which then allows to search for the resulting visible decay products. We look forward to an implementation of this search strategy in present and upcoming runs at the LHC to explore new avenues in the hunt for DM. 

\acknowledgments

We thank Ulrich Haisch for helpful discussions. This work is supported by the German Science Foundation
(DFG) under the Collaborative Research Center~(SFB) 676 Particles,
Strings and the Early Universe as well as the ERC Starting Grant `NewAve'
(638528). The work of BP is funded by the STFC UK. We are also grateful for the travel support for BP to visit DESY provided by the PIER Fellowship PFS-2016-14. PIER (Partnership for Innovation, Education and Research) is the strategic partnership between DESY and Universit{\"a}t Hamburg.

\flushbottom

\providecommand{\href}[2]{#2}\begingroup\raggedright\endgroup

\end{document}